\documentstyle[sprocl]{article}

\input{psfig}
\def\Journal#1#2#3#4{{#1} {\bf #2}, #3 (#4)}

\def\NPB{{\em Nucl. Phys.} B}
\def\PLB{{\em Phys. Lett.}  B}

\def\PRD{{\em Phys. Rev.} D}

\def\be{\begin{equation}}
\def\ee{\end{equation}}
\def\bea{\begin{eqnarray}}
\def\eea{\end{eqnarray}}
\begin{document}

\rightline{{\bf IEM--FT--117/95}}
\rightline{{\bf OUTP--95--35P}}
\vskip0.8cm

\title{SUPERGRAVITY AND THE $b \rightarrow s,\gamma$ DECAY}

\author{B. DE CARLOS }

\address{Department of Theoretical Physics, 1 Keble Road, \\
Oxford OX1 3NP, UK}

\author{J.A. CASAS }

\address{Instituto de Estructura de la Materia (CSIC), Serrano 123, \\
28006 Madrid, Spain}

%%%%%%%%%%%%%%%%%%%%%%%%%%%%%%%%%%%%%%%%%%%%%%%%%%%%%%%%%%%%%%
% You may repeat \author \address as often as necessary      %
%%%%%%%%%%%%%%%%%%%%%%%%%%%%%%%%%%%%%%%%%%%%%%%%%%%%%%%%%%%%%%

\maketitle\abstracts{
We evaluate the branching ratio BR($b\rightarrow s,\gamma$) in the
minimal supersymmetric standard model (MSSM), determining the
corresponding phenomenological restrictions on two attractive
supergravity scenarios, namely minimal supergravity and a class of
models with a natural solution to the $\mu$ problem. We have included
in the calculation some one--loop refinements that have a substantial
impact on the results.
It is stressed the fact that
an eventual improvement of the experimental bounds of order $10^{-4}$
would strengthen the restrictions on the MSSM dramatically. This would
be enough to discard these supergravity scenarios with $\mu<0$ if
no discrepancy is found with the standard model prediction, while for
$\mu>0$ there will remain low-energy windows.}

\section{The $b \rightarrow s, \gamma$ decay in SUGRA models}

The new experimental data available\cite{baris94} place stringent
upper and lower bounds on the branching ratio (BR) for the process
$b\rightarrow s,\gamma$:
$1\times 10^{-4}<$ BR($b\rightarrow s,\gamma$) $<4\times 10^{-4}$ .
On the other hand it is well known that this process has the
potential to put relevant constraints to physics beyond the Standard
Model (SM).
In particular, we want to study the prediction that for this BR give
two different supergravity (SUGRA) models whose particle content at
low energies is that of the Minimal Supersymmetric Standard Model
(MSSM)\footnote{A detailed description of the calculation is given
elsewhere~\cite{decar95}.}. That is, we will evaluate the expression
(in units of the BR for the semileptonic $b$ decay):
\begin{eqnarray}
\frac{BR(b \rightarrow s \gamma)}{BR(b \rightarrow c e \bar{\nu})} & = &
\frac{|K_{ts}^{*} K_{tb}|^{2}}{|K_{cb}|^{2}} \frac{6 \alpha_{QED}}{\pi}
\nonumber \\
& \times & \frac{\left [\eta^{16/23} A_{\gamma} + \frac{8}{3}
(\eta^{14/23} - \eta^{16/23}) A_{g} + C \right]^{2}}{I(z)} F
\label{BR}
\end{eqnarray}
where $z=\frac{m_c}{m_b}$, $\eta=\frac{\alpha_s(M_W)}{\alpha_s(m_b)}$,
$I(z)$ is the phase space factor, $C$ stands for the leading
logarithmic QCD corrections, and $F$ contains NLO effects. Finally,
$A_{\gamma,g}= A_{\gamma,g}^{SM} +
A_{\gamma,g}^{H^{-}}+A_{\gamma,g}^{\chi^{-}}$
are the coefficients of the effective operators for the $bs\gamma$ and
$bsg$ interactions\cite{berto91}; in our case we consider as relevant
the contributions coming from the SM diagram (top quark and $W^{-}$)
plus those with top quark and charged Higgs, and stops/scharms and
charginos running in the loop. It is interesting to note that both
$A_{\gamma,g}^{SM}$ and $A_{\gamma,g}^{H^{-}}$ have always the same
sign, giving therefore a total amplitude which is bigger than the SM
one. However the presence of the chargino contribution which, for a
wide range of the parameter space has opposite sign to that of the
other two amplitudes, softens this effect. This is not enough, in
general, to lower the total BR below the SM prediction, although in
some cases it might be possible to have such big values for
$A_{\gamma,g}^{\chi^{-}}$ as to drive the total BR even below the CLEO
lower bound.

The MSSM is defined by the superpotential, $W$, and the form of the
soft supersymmetry breaking terms. $W$ is given by
\begin{equation}
\label{W}
W=\sum_i\left\{
h_{u_i}Q_i H_2 u_i^c + h_{d_i}Q_i H_1 d_i^c
+ h_{e_i}L_i H_1 e_i^c \right\} +  \mu H_1 H_2\;\; ,
\end{equation}
where $i$ is a generation index, $Q_i$ ($L_i$) are the scalar partners
of the quark (lepton) SU(2) doublets, $u^c_i,d^c_i$ ($e^c_i$) are the
quark (lepton) singlets and $H_{1,2}$ are the two SUSY Higgs doublets;
the $h$--factors are the Yukawa couplings and $\mu$ is the usual Higgs
mixing parameter. The soft breaking terms have the form
\begin{eqnarray}
\label{Lsoft}
{\cal L}_{\rm soft}&=&\frac{1}{2}M\lambda_a\lambda_a -
\sum_j m^2 |\phi_j|^2\ -\ \sum_i A\left[
h_{u_i}Q_i H_2 u_i^c + h_{d_i}Q_i H_1 d_i^c
\right.
\nonumber \\
&+&  \left. h_{e_i}L_i H_1 e_i^c + {\rm h.c.} \right]
- \left[ B\mu H_1 H_2 + {\rm h.c.}\right] \;\; ,
\end{eqnarray}
where $a$ is gauge group index and $\lambda_a$ are the gauginos.
$m$, $M$, $A$, $B$ are the scalar and gaugino masses and the
coefficients of the trilinear and bilinear scalar terms,
respectively. They are assumed to be universal at the unification
scale $M_X$.

The aim of our calculation is to obtain the restrictions on the MSSM
from the $b\rightarrow s,\gamma$ decay in the two SUGRA scenarios
defined by:

\noindent {\bf SUGRA I}

\noindent This is just the minimal SUGRA theory. It is defined by a
K\"ahler potential $K=\sum_j|\phi_j|^2$ and a gauge kinetic function
$f_{ab}=\delta_{ab}$, so that all the kinetic terms are canonical,
whereas the superpotential $W$ is assumed to be as in eq.~(\ref{W}),
$\mu$ being an inicial parameter. Then all the soft terms are
automatically universal and the coefficients $A$ and $B$ are related
to each other by the well known relation $B=A-m$. This scenario has a
serious drawback, namely the unnaturally small (electroweak) size of
the initial $\mu$ parameter in the superpotential (often known as
$\mu$ problem); that leads us to the next scenario.

\noindent {\bf SUGRA II}

\noindent Recently, there have appeared several attractive mechanisms
to solve the $\mu$ problem~\cite{casas93} that, quite remarkably,
lead, in the presence of a  K\"ahler potential as in minimal SUGRA, to
a similar prediction for the value of $B$, namely $B=2m$. This can be
achieved also if, as suggested by many superstring constructions, the
kinetic terms are non-universal~\cite{kaplu93}. From now on we will
refer (any of) these scenarios as SUGRA II. The corresponding MSSM
emerging from them is as in minimal SUGRA except for the value of $B$.

So our starting point is the MSSM with initial parameters
$\alpha_X,M_X,h_t,h_b,$\newline$h_{\tau},\mu,m,M,A,B$.
By consideration of one of the two previous SUGRA theories we
eliminate the $B$ parameter, and then we demand consistency with all
the experimental data, that is:

{\em i)} Correct unification of the gauge coupling
constants

\noindent This can only be achieved when the renormalization group
equations (RGE) of the gauge couplings are taken at two--loop order.
For consistency, all the supersymmetric thresholds (and the top quark
one) have to be taken into account in the running in a separate way.
Therefore, given a choice of the initial parameters, an iterative
process is necessary to achieve an agreement between the resulting
prediction for $\alpha_1(M_Z),\alpha_2(M_Z),\alpha_3(M_Z)$ and their
experimental values~\cite{amald91}.

{\em ii)} Correct masses for all the observed particles.

\noindent Concerning the masses of the fermions, the boundary
conditions for Yukawa couplings of the top and bottom quarks and the
tau lepton, $h_t,h_b,h_\tau$, have to be chosen so that the
experimental masses are properly fitted (note that the running masses
$m_{t,b,\tau}(Q)$ do not coincide with the physical (pole) masses
$M_{t,b,\tau}$). Notice also that, as mentioned above, the masses in
the BR($b\rightarrow s,\gamma$) are the running masses at the
electroweak scale.

{\em iii)} Masses for the unobserved particles compatible with
the experimental bounds~\cite{pdg}.

{\em iv)} Correct electroweak breaking, i.e. $M_Z=M_Z^{exp}$.

\begin{figure}
\centerline{
%% FOLLOWING LINE CANNOT BE BROKEN BEFORE 80 CHAR
\psfig{figure=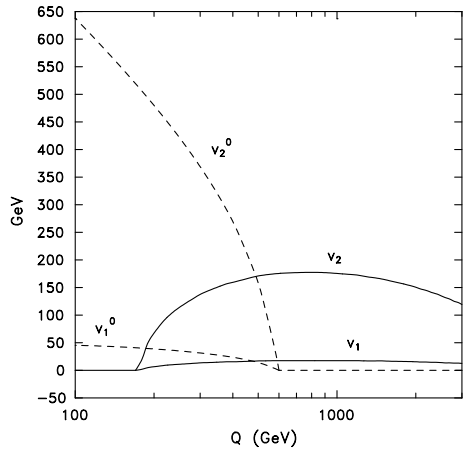,height=6.0cm,bbllx=3.cm,bblly=3.5cm,bburx=18cm,bbury=16cm}}
\caption{Plot of $v_1\equiv\langle H_1\rangle$, $v_2\equiv\langle H_2\rangle$
vs the $Q$ scale between 100 GeV and 3 TeV for the
model defined by $m=M=300$ GeV,
$A=317$ GeV, $B=A-m$, $\mu=403.76$ GeV, $h_t=0.568$, $h_b=0.063$,
$h_\tau=0.072$, $\alpha_X=0.0404$ (all
quantities defined at $M_X=1.6\times 10^{16}$ GeV).
Solid lines: complete one--loop results,
dashed lines: (renormalization improved) tree level results.
}
\end{figure}

\noindent The vacuum expectation values (VEVs) of the two Higgses,
$v_1=\langle H_1 \rangle$, $v_2=\langle H_2 \rangle$ (upon which $M_Z$
depends) are to be obtained from the minimization of the Higgs
potential. The tree level part of this in the MSSM has the form
$V_o=m_1^2 |H_1|^2 + m_2^2 |H_2|^2 + 2\mu B H_1H_2
+ \frac{1}{8}(g^2+{g'}^2)(|H_2|^2-|H_1|^2)^2$,
where all the parameters are understood to be running parameters
evaluated at the renormalization scale $Q$. By a suitable redefinition
of the phases of the fields it is always possible to impose
$v_1,v_2>0$. As was clarified some time ago~\cite{gambe90}, $V_o$ and
the corresponding tree level VEVs $v_1^o(Q),v_2^o(Q)$ are strongly
$Q$--dependent. In order to get a much more scale independent
potential the one--loop correction $\Delta V_1$ is needed. This is
given by
\begin{equation}
\label{DeltaV1}
\Delta V_1={\displaystyle\sum_{j}}{\displaystyle\frac{n_j}{64\pi^2}}
M_j^4\left[\log{\displaystyle\frac{M_j^2}{Q^2}}
-\frac{3}{2}\right]\;\;,
\end{equation}
where $M_j^2(\phi,t)$ are the tree-level (field--dependent) mass
eigenstates and $n_j$ are spin factors. In this way, the minimization
of $V=V_o+\Delta V_1$ gives one-loop VEVs $v_1(Q),v_2(Q)$ much more
stable against variations of the $Q$ scale. In general, there is a
region of $Q$ where $v_1(Q),v_2(Q)$ are remarkably $Q$--stable and a
particular scale, $\hat{Q}$, always belonging to that region, at which
$v_1(\hat{Q}),v_2(\hat{Q})$ essentially coincide with
$v_1^o(\hat{Q}),v_2^o(\hat{Q})$. This is illustrated in Fig.~1.

Here we have evaluated $v_1,v_2$ at $\hat{Q}$ and then obtained
$v_1(Q)$, $v_2(Q)$ at any scale by using the RG running of the $H_1$,
$H_2$ wave functions. Furthermore, we have included in
eq.~(\ref{DeltaV1}) {\em all} the supersymmetric spectrum since some
approximations can lead to wrong results~\cite{decar93}.

Conditions {\em (i)--(iii)} allow to eliminate
$\alpha_X,M_X,h_t,h_b,h_{\tau}$, while {\em (iv)} allows to eliminate
one of the remaining parameters, which we choose to be  $\mu$. This
finally leaves three independent parameters, namely $m$, $M$ and $A$.
Actually, for many choices of these parameters there is no value of
$\mu$ capable of producing the correct electroweak breaking, a feature
which restricts the parameter space substantially. On the other hand,
there can be two branches of solutions depending on the sign of $\mu$.

\begin{figure}
\centerline{
\psfig{figure=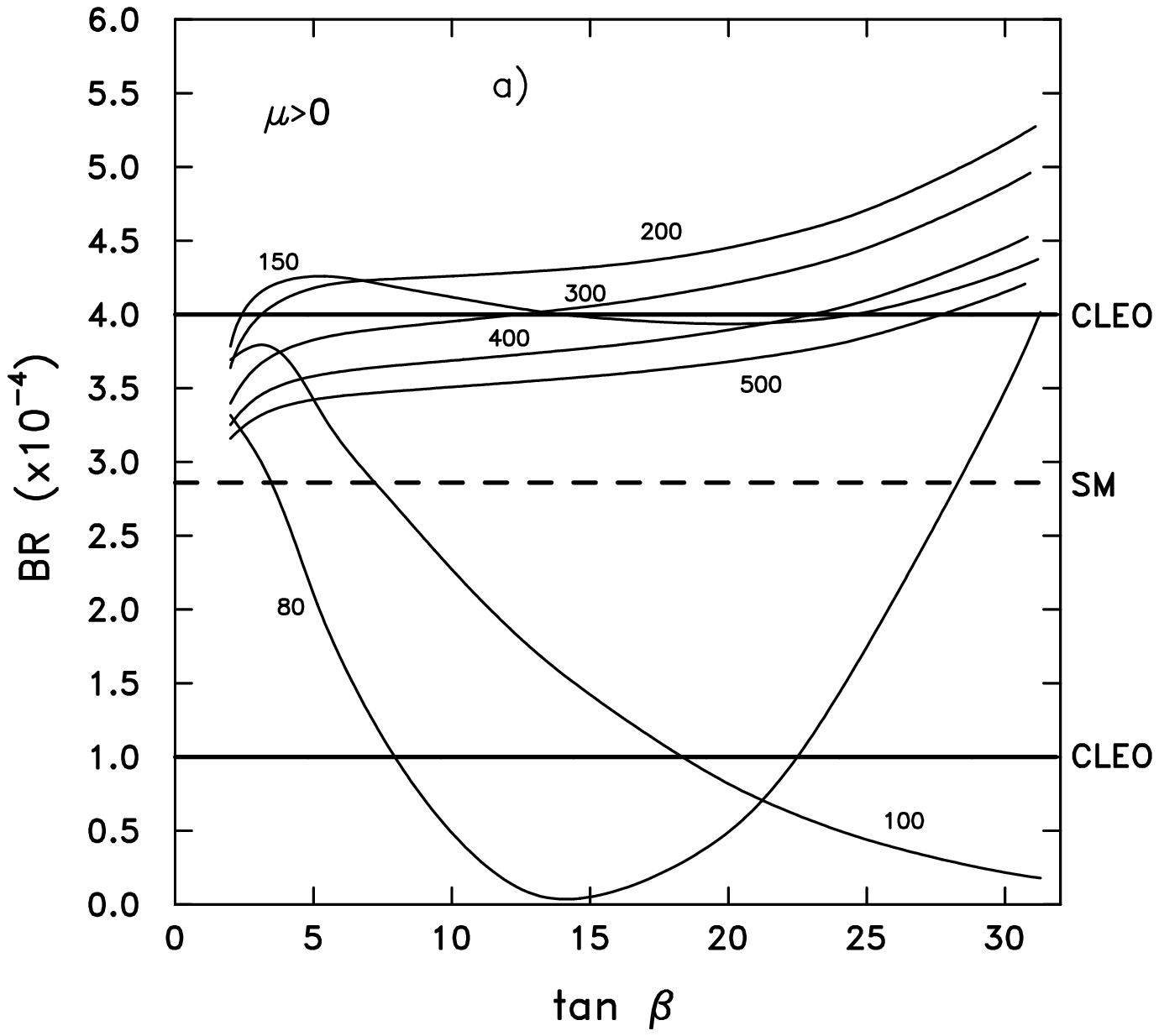,height=5cm,bbllx=3cm,bblly=3.5cm,bburx=18cm,bbury=16cm}
\psfig{figure=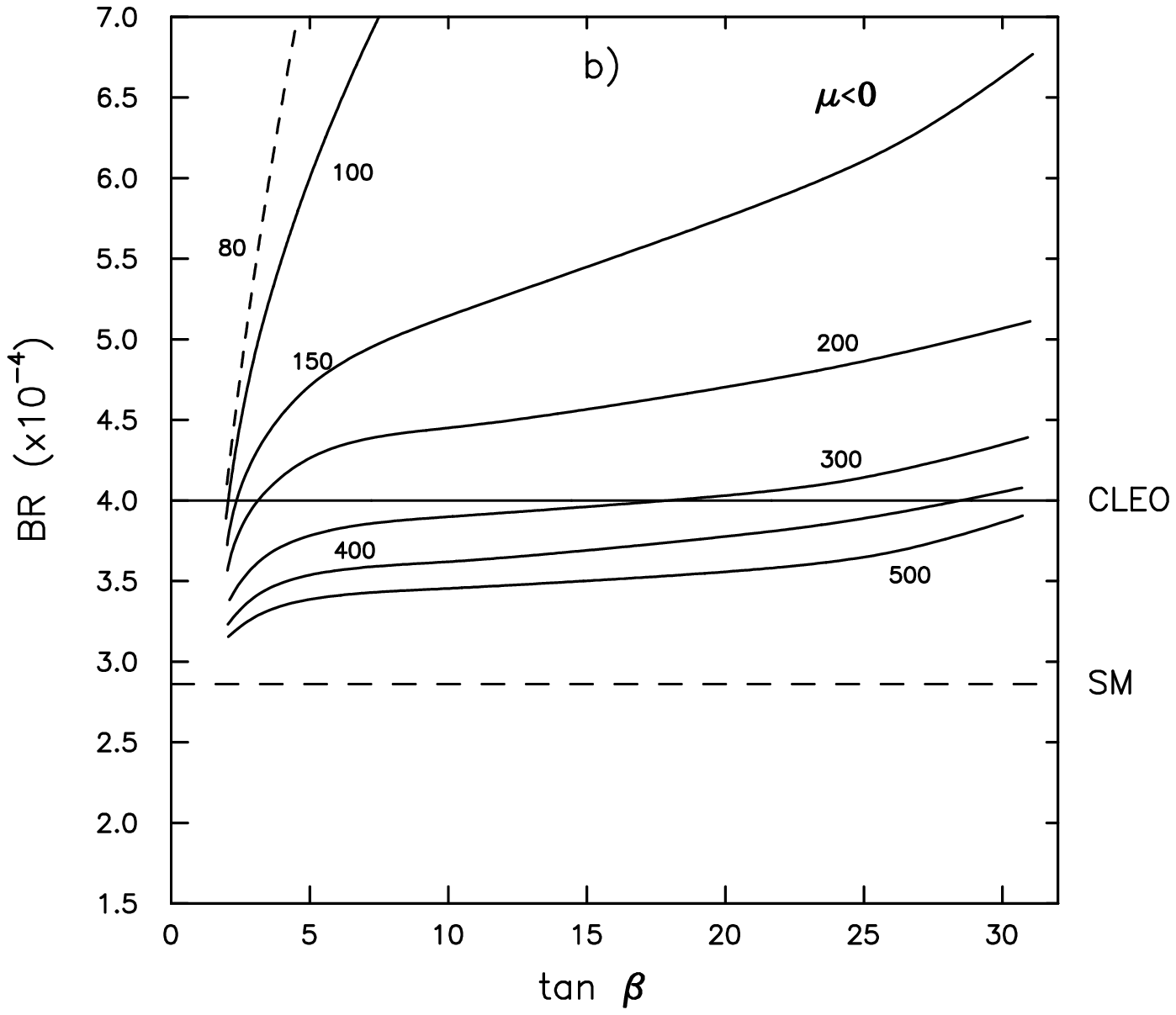,height=5cm,bbllx=3cm,bblly=3.5cm,bburx=18cm,bbury=16cm}}
\caption{Plot of the branching ratio
BR($b\rightarrow s,\gamma$) vs
$\tan \beta$ for the SUGRA I scenario (minimal SUGRA) with $m=M$ for
different
values of $m$ (80, 100, 150, 200, 300, 400, 500 GeV)
: a) branch $\mu>0$; b) branch $\mu<0$. The dashed curves indicate that
the model becomes incompatible with the experimental lower bounds on
supersymmetric particles.
The Standard Model
prediction (SM) and the CLEO bounds are also shown in the figure.
}
\end{figure}

\section{Results}

\noindent {\bf SUGRA I}

\noindent As discussed above, our parameter space is now restricted to
$m,M,A$. In order to present the results in a comprehensible way, let
us make the assumption $m=M$, and trade the high energy parameter $A$
by the low energy one $\tan \beta=v_2/v_1$. The corresponding plots of
the branching ratio BR($b\rightarrow s,\gamma$) versus the remaining
parameter, $\tan \beta$ for different values of $m$ are given in
Fig.~2a (branch $\mu>0$) and Fig.~2b (branch $\mu<0$). We can see that
for $m\geq 200$ GeV and both $\mu>0$, $\mu<0$, for each value of $m$
there is a maximum acceptable value for $\tan \beta$. In general for
$m\geq 200$ GeV  the restrictions are stronger for positive values of
$\mu$ and small values of $m$. For $m\leq 200$ GeV, however, the
situation is different: whereas for $\mu<0$ the restrictions become
very strong (only a narrow range of $\tan \beta$ is allowed), for $\mu>0$
there appear large windows of allowed values of $\tan \beta$. Here the
CLEO lower bound plays a relevant role.

It is clear that the trend to the SM result is so slow that an
improvement of the CLEO bounds (particularly the upper one) on BR of
order $10^{-4}$ would push the lower limits on $m$ to the TeV region
(except for the above--mentioned windows in the $\mu>0$ case).

\noindent {\bf SUGRA II}

\begin{figure}
\centerline{
\psfig{figure=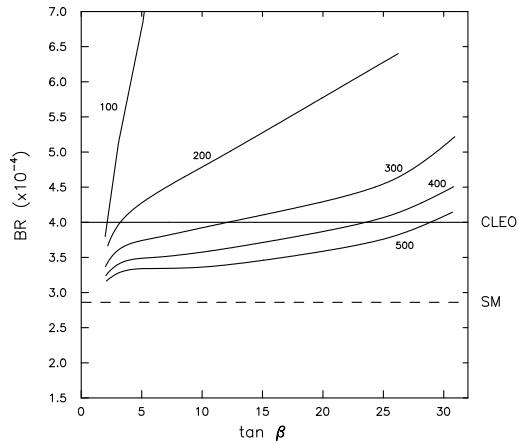,height=6.cm,bbllx=3cm,bblly=3.5cm,bburx=18cm,bbury=16cm}}
\caption{The same as in Fig.~2, but for the SUGRA II scenario.}
\end{figure}

\noindent We note here that the sign of $\mu$ is always negative since
(adopting the convention $v_1,v_2 > 0$) for $\mu>0$ and $B=2m$ the
necessary electroweak breaking cannot be achieved. The SUGRA II
results present a similar pattern to those of SUGRA I with $\mu<0$.
Again, we find from Fig.~3 that for a given value of $m$ there is a
maximum acceptable value for $\tan \beta$. The bound becomes less
stringent as $m$ increases; therefore, once again, an improvement of
the CLEO bounds (especially the upper one) of order $10^{-4}$ would
amount to a dramatical improvement of the MSSM constraints.

\section*{Acknowledgements}

The speaker (BdeC) wants to thank the Organizing Committee of this
workshop for their kind hospitality. This work has been partly
supported  by CICYT under contract AEN94-0928, and by the European
Union under contract No. CHRX-CT92-0004. BdeC was supported by a
Spanish MEC Postdoctoral Fellowship.

\end{document}